\documentclass[ amsmath, amssymb, 12pt]{article}

\usepackage{amsmath, amssymb}
\usepackage{textcomp}
\usepackage[space]{cite}
\usepackage{setspace}
\begin{document}


\def\al{\alpha}
\def\d{\partial}
\def\t0{\tilde{0}}
\def\ta{\tilde{a}}
\def\tb{\tilde{b}}
\def\tc{\tilde{c}}
\def\td{\tilde{d}}
\newcommand{\be}{\begin{equation}}
\newcommand{\ee}{\end{equation}}
\newcommand{\bea}{\begin{eqnarray}}
\newcommand{\eea}{\end{eqnarray}}
\newcommand{\talpha}{\tilde{\alpha}}


\begin{flushright}

\end{flushright}
\vspace{10pt}
\begin{center}
  {\LARGE \bf Self-accelerating Universe in modified gravity with dynamical torsion} \\
\vspace{20pt}
V.~Nikiforova$^a$, S.~Randjbar-Daemi$^b$,
V.~Rubakov$^{a,c}$\\
\vspace{15pt}
  $^a$\textit{Institute for Nuclear Research of
         the Russian Academy of Sciences,\\  60th October Anniversary
  Prospect 7a, 117312 Moscow, Russia
  }\\
\vspace{5pt}
$^b$\textit{The Abdus Salam International Centre for Theoretical Physics, \\ Strada Costiera 11 34014, Trieste, Italy
}\\
\vspace{5pt}
$^c$\textit{Department of Particle Physics and Cosmology,\\
Physics Faculty, Moscow State University\\ Vorobjevy Gory,
119991, Moscow, Russia
}
    \end{center}
    \vspace{5pt}


\begin{abstract}
We consider a model belonging to the class of gravities with dynamical torsion. The model is free of ghosts and gradient instabilities about Minkowski and torsionless Einstein backgrounds. We find that at zero cosmological constant, the model admits a self-accelerating solution with non-Riemannian connection. Small value of the effective cosmological constant is obtained at the expense of the hierarchy between the dimensionless couplings.
\end{abstract}


\section{Introduction}
The possibility of modifying General Relativity at large distances is both theoretically exciting and potentially interesting for cosmology. In particular, IR-modified gravity may serve as an explanation of the accelerated late-time expansion of the Universe, alternative to the cosmological constant and dark energy. However, it often happens that self-accelerating solutions in IR-modified gravities are plagued with instabilities. A famous example is the DGP model \cite{9}, which admits both Minkowski and self-accelerating backgrounds \cite{41}. The latter, however, has ghost instability \cite{27}.

There are various approaches to modifying gravity in IR, see reviews \cite{14, 37, 21, 36, 39, 35}. Particularly promising candidates for consistent modified gravities are theories with dynamical torsion. These are often considered in the framework of Poincar\'e gauge gravities \cite{book1, book2, book3}, whose very general versions are now available \cite{H, H2}. A defining property of theories with dynamical torsion is that they treat the vierbein and connection as independent variables. Therefore, the spectrum of linearized perturbations in these theories is generically reacher than in General Relativity, and often contains massive excitations. This is precisely the reason for expecting that gravity is modified in IR.

It is natural to ask whether models with dynamical torsion admit stable self-accelerating cosmological solutions, without explicit cosmological constant term in the action. Solutions of this sort were indeed found in various Poincar\'e gravity models \cite{Hehl4, 42, 49, 44, 50, 47, 51, 46, 48}. The key issue then is the stability of the self-accelerating solutions: as mentioned above, in other IR modified gravities self-acceleration is often unstable because of the ghost and/or gradient instabilities in the spectrum of linearized perturbations.

Having in mind these stability issues, one is led to search for self-accelerating solution(s) in models which are stable in {\it Minkowski} background. Several classes of models with the latter property were found in Refs.~\cite{29, 30, 31, 32}. One of these models was further studied in Refs.~\cite{33, 34, 45} where it was shown that ghosts, gradient instabilities and tachyons are absent also in de Sitter and anti-de Sitter spaces and arbitrary torsionless Einstein backgrounds of sufficiently small curvature. The gravitational interaction is mediated by both massless and massive spin-2 fields, with relative strength being a free parameter \cite{33}. The simplicity and nice properties of this particular model make it worth exploring further. To the best of our knowledge, it is not known whether other theories with dynamical torsion, including those with the Holst term \cite{Holst} and more general actions \cite{H, H2}, have similar stability properties about various backgrounds. 

The purpose of this and forthcoming papers is to study whether the model of Refs.~\cite{33, 34, 45} without explicit cosmological constant term in the action admits a self-accelerating solution, and if so, whether this solution is stable. The notion of accelerating background needs qualification, however, since the behavior of matter in a non-trivial background depends on its interaction with metric and connection fields. Unlike in Poincar\'e gravities, we assume here that matter interacts with gravity exactly like in General Relativity, then the acceleration is equivalent to de Sitter-like behavior of metric, irrespectively of the properties of the background connection.

Another option would be to follow the Poincar\'e gravity route which leads to direct interaction of connection field with the spin of matter. As discussed in Ref.~\cite{49}, the effect of this interaction on cosmological solutions is small at ordinary matter densities (but may strongly influence the dynamics in the early Universe \cite{Hehl3}), so the solution  we are after is not expected to get grossly modified. However, non-trivial background connection affects the propagation of particles with spin, giving rise to strong constraints on the cosmic connection (see Ref.~\cite{obs_constraints} and references therein). This is our main motivation to deviate from the Poincar\'e gravity framework and consider the connection merely as an extra field interacting with vierbein only.

In this paper we show that the model of Refs.~\cite{33, 34, 45} does admit a self-accelerating solution under mild assumptions on its couplings. The key issue is then the stability of perturbations about this solution. We plan to address this issue in future publications.

The paper is organized as follows. We present the Lagrangian and remind the earlier results in Section 2. In Section 3 we study spatially flat homogenous and isotropic cosmology and find that the model admits a self-accelerating solution with de-Sitter metric and non-zero torsion. In Section 4 we demonstrate that the effective dark energy density can be made small by an appropriate choice of parameters.

\section{The Model}
We make use of the tetrad formalism. The vierbein and connection are considered as independent fields. Following the notations of Refs. \cite{33, 34, 45}, we denote the vierbein by $e^i_\mu$ and connection by $A_{ij\mu} = -A_{ji\mu}$, where $\mu = (0,1,2,3)$ is the space-time index, and $i,j = (0,1,2,3)$ are the tangent space indices. The latter are raised and lowered using the Minkowski metric $\eta_{ij}$, so we do not distinguish upper and lower tangent space indices in what follows, if this does not lead to an ambiguity. The signature of metric is $(-, +, +, +)$.

The action of the model is \cite{34}:
\be  S= \int~d^4x~eL  \; , \;\;\; L= \frac{3}{2} ( \tilde{\alpha} F -
\alpha R) +
 c_3 F^{ij}F_{ij} + c_4 F^{ij}F_{ji} + c_5 F^2 + c_6 (\epsilon \cdot F)^2 \; , \label{all_2} \ee
 where
 $ e \equiv det(e^i_\mu) \; ; $
$F_{ijkl}$ is the curvature tensor constructed with the connection $A_{ij\mu}$,
\[
F_{ijkl} = e^\mu_k e^\nu_l ( \partial_\mu A_{ij\nu} - \partial_\nu A_{ij\mu} + A_{im\mu}A_{m j\nu} - A_{jm\mu}A_{m i\nu} ) \; ; 
\]
\[
F_{ij}=\eta^{kl}F_{ikjl}\;,\;\; F=\eta^{ij}F_{ij} \; , \;\; \epsilon \cdot F \equiv \epsilon^{ijkl}F_{ijkl} 
\; ;
\]
$\epsilon_{ijkl}$ is the Levi-Civita symbol defined in such a way that $\epsilon^{0123} = - \epsilon_{0123} = 1$;
$R_{ijkl}$ is the Riemannian curvature tensor,
\[ R_{ijkl}= e^\mu_k e^\nu_l ( \partial_\mu \omega_{ij\nu} - \partial_\nu
\omega_{ij\mu} + \omega_{im\mu}\omega_{mj\nu} - \omega_{jm\mu}\omega_{mi\nu} )\; ;
\]
\[ R_{ij}=\eta^{kl}R_{ikjl}\;,\;\; R=\eta^{ij}R_{ij}\; , \]
where $\omega_{ij\mu}$ is the Riemannian spin-connection. It is expressed in terms of the vierbein as follows:
\[
 \omega_{ij\mu} \equiv \omega_{ijk}e^k_\mu =\frac{1}{2}( C_{ijk}-C_{jik} - C_{kij} )e^k_\mu \; ,  
\]
where
\[
C_ {ijk} = e_j^\mu e_k^\nu (\d_\mu e_{i\nu} - \d_\nu e_{i\mu}) \; . 
\]
The constants $ \alpha, \tilde{\alpha}, c_3, c_4, c_5, c_6$ are the parameters. We impose the following conditions,
\begin{subequations}
\begin{align}
& c_3+c_4=-3c_5 \;, \label{all_19} \\
&\alpha < 0, \;\;\;\; \tilde{\alpha} > 0, \;\;\;\; c_5 < 0, \;\;\;\; c_6 > 0  \label{all_13} \; , \end{align} \label{s1}
\end{subequations}
in order not to have the pathological degrees of freedom in the Minkowski background \cite{31}. Finally, we impose yet another condition,
\be  c_5 + 16 c_6 < 0 \;.  \label{all_62} \ee
We will see in Sec.~\ref{sec:background solution} that the latter condition ensures the existence of the self-accelerating solution.

It is worth noting that the action \eqref{all_2} is equivalent to the action used in \cite{29, 30, 31, 33, 45}, which, instead of the explicit $\al R$-term, involves mass terms for torsion. To this end, one decomposes the connection as follows,
\[
 A_{ij\mu} \equiv A_{ijk}e^k_\mu =\left[\frac{1}{2} \left(T_{ijk}-T_{jik} - T_{kij} \right)+ \omega_{ijk} \right]e^k_\mu \; ,  
\]
where $T_{ijk}=-T_{ikj}$ is the torsion tensor. The latter can be decomposed into its irreducible components under the local $O(1,3)$ group,
\[
T_{ijk}=\frac{2}{3}(t_{ijk}-t_{ikj}) + \frac{1}{3}(\eta_{ij}v_k - \eta_{ik}v_{j}) + \epsilon_{ijkl}a^l \; ,
\]
where the field $t_{ijk}$ is symmetric with respect to the interchange of $i$ and $j$ and satisfies the cyclic and trace identities,
\[
t_{ijk}+t_{jki}+t_{kij}=0, \qquad \eta^{ij}t_{ijk}=0, \qquad \eta^{ik}t_{ijk}=0 \; .
\]
The action is then
\[
S=\int~d^4x~e L \; ,
\]
where
\[
L = \frac{3}{2}(\talpha - \al)F + \al(t_{ijk}t^{ijk} - v_iv^i + \frac{9}{4}a_ia^i) + c_3F^{ij}F_{ij} + c_4F^{ij}F_{ji} + c_5F^2 + c_6(\epsilon \cdot F)^2 \; .
\]
We will use the action \eqref{all_2} in what follows.

In Ref. \cite{33} it was found that there are three propagating modes at the linear level in the Minkowski background: the massless spin-2 mode, the massive spin-2 mode with mass
\be m^2=\frac{\talpha(\tilde{\alpha}-\alpha)}{2\alpha c_5} \label{all_80} \ee
 and the massive spin-0 mode with mass \be m^2_0=\frac{\tilde{\alpha}}{16c_6} \;. \label{all_81} \ee There are no ghosts or tachyons in the Minkowski background. In the theory equipped with the cosmological constant, the perturbations are healthy in torsionless Einstein backgrounds of sufficiently small curvature as well.

The analysis of gravitational interactions between sources in Minkowski background \cite{33} reveals that for small $m$ there is the correspondence,
\be  \al = - \frac{M_{Pl}^2}{24\pi} \;.  \label{all_83} \ee
We leave other parameters arbitrary for the time being.

\section{\label{sec:background solution}The self-accelerating solution}
We consider now spatially flat homogeneous and isotropic cosmology. Homogeneity and isotropy dictate the following most general Ansatz:
\be
e^{\t0}_0 = N(t) \;,\;\;\;\; e^{\ta }_b = a(t) \delta^{\ta}_b
 \;,\quad
A_{\t0 \ta \tb}=f(t)\delta_{\ta\tb}  \;, \quad
A_{\ta\tb\tc}=g(t)\varepsilon_{\ta\tb\tc}\;,
\label{c1}
\ee
and the remaining components of metric and connection vanish. Here $a, \tilde{a}=(1,2,3)$, tilde denotes tangent space indices, while space-time indices do not have tilde. In other words, $i=(\t0 , \ta)$, $\mu = (0,a)$. Note that due to the antisymmetry of $A_{ijk}$ with respect  to the  interchange of the first pair of indices no other components of connection can be non-vanishing.

We shall show that the Ansatz 
\be
a=e^{\lambda t} \;, \qquad \lambda=const\;,   \label{dS}
\ee
will necessarily imply that
$$ f=const, \qquad g=const \; .$$

With the Ansatz \eqref{c1}, we calculate the non-vanishing components of the curvature tensor
$F_{ijkl}$,
$$
F_{\t0 \ta \t0 \tb}= \frac{1}{Na} \partial_0 (af) \delta_{ab} \; ,
\quad F_{\t0\ta\tb\tc}= -2 fg \varepsilon _{abc} \;,
$$
\[
F_{\ta\tb\t0\tc}= \frac{1}{Na} \partial_0(ag) \varepsilon_{abc}
\; , \quad
F_{\ta\tb\tc\td}= (f^2-g^2)(\delta_{ac}\delta_{bd} - \delta_{ad}\delta_{bc})\;.
\]
The non-vanishing components of $F_{ij}$ are
\[
F_{\t0\t0}= \frac{3}{Na}\partial_0(a f)
\; , \quad
F_{\ta\tb}= \left[2(f^2-g^2)-\frac{1}{Na}\partial_0(af)\right]\delta_{ab} \;.
\]
Note that $F_{ij}$ is symmetric as required by the homogeneity.
Finally, we calculate $F=\eta_{ij}F_{ij}$, $\epsilon \cdot F$ and $R \equiv R_{ijkl}\eta^{ik}\eta^{jl}$:
\begin{align*}
& F= 6\left[f^2-g^2-\frac{1}{Na}\partial_0(af)\right] \; ,
 \quad
\epsilon \cdot F = 12\left[ -2fg+\frac{1}{Na}\partial_0(ag)  \right] \; ,
\\
& R = 6 \left[ \frac{1}{Na}\partial_0\left(\frac{\dot{a}}{N}\right) + \frac{1}{a^2}\left(\frac{\dot{a}}{N}\right)^2 \right] \;.
\end{align*}
The action (\ref{all_2}) in terms of the homogeneous and isotropic fields is
\begin{align*}
 eL=& 9\talpha \left[Na^3 ( f^2 -g^2) - a^2\partial_0 (af)\right]- 9\al \left[ a^2\d_0 \left(\frac{\dot{a}}{N}  \right) + \frac{a}{N} (\dot{a})^2 \right] \nonumber \\
&-  36 c_5 a^2 (f^2-g^2) \partial_0(a f)+ 144 c_6\left\{ 4Na^3 f^2g^2  - 4a^2 fg \partial_0(ag )
+ \frac{a}{N} \left[\partial_0 (ag) \right]^2 \right\}  \;,
\end{align*}
where $\dot{a} \equiv \d_0a$.
Note that, due to \eqref{all_19} and symmetry of $F_{ij}$, this expression contains only $c_5$, $c_6$, $\al$ and $\talpha$, but not $c_3$ and $c_4$ separately. The terms with $(f^2 - g^2)^2$ and $[\partial_0 (af)]^2$ have canceled out also due to \eqref{all_19}. Upon integrating by parts we write the action in the following form:
\begin{align}
eL=&9\talpha\left[(f^2-g^2)Na^3+2a^2\dot{a}f\right]+ 9\al \frac{a}{N}(\dot{a})^2+36c_5g^2a^2\d_0(af)  \nonumber \\
&+144c_6\left\{4f^2g^2Na^3-4a^2fg\d_0(ag)+\frac{a}{N}[\d_0(ag)]^2\right\} \;. \label{all_4}
\end{align}

There are three independent equations of motion that follow from (\ref{all_4}).
We choose the gauge $N=1$ after varying with respect to $N$, $f$ and $g$, divide by
$a^3$ and obtain
\begin{subequations}
\bea
\frac{\delta}{\delta N}&:& \;\;\;\;
\talpha(f^2 - g^2) - \al \frac{\dot{a}^2}{a^2}
- d_6 \frac{[\partial_0(ag)]^2}{a^2} + 4d_6 f^2 g^2 = 0 \;, \label{all_27}
\\
\frac{\delta}{\delta f}&:& \;\;\;\;
 \talpha(f+\frac{\dot{a}}{a})
- d_5 g \frac{\partial_0 (ag)}{a} + 4d_6 fg^2 = 0  \;, \label{all_28}
\\
\frac{\delta}{\delta g}&:& \;\;\;\;
-\talpha g + d_5 g \frac{\partial_0 (af)}{a}
- d_6 \frac{\partial_0 [a\partial_0 (ag)]}{a^2} + 4d_6 f^2 g = 0 \;,  \label{all_29}
\eea
\end{subequations}
where we have introduced the notations
$$ d_6 \equiv 16c_6\;, \;\; d_5 \equiv 4c_5+32c_6\;. $$
We are interested in a self-accelerating solution of this system with the scale factor given by eq.~\eqref{dS}. Such a solution necessarily has time-independent $f$ and $g$. To see this, we note that Eq.~\eqref{all_28} can be solved for $\d_0 g$ as function of $f$ and $g$. Then Eq.~\eqref{all_27} becomes an algebraic equation that determines $f=f(g)$. Hence, we can express $\d_0g$, $\d^2_0g$ and $\d_0f$ as algebraic functions of $g$. Plugging these into Eq.~\eqref{all_29} we obtain an algebraic equation for $g$ with time-independent coefficients. We have checked that $g$ does not drop out of the latter equation (which, in fact, can be cast into the eighth order equation for $g^2$). The solution to that equation is independent of time, as claimed.

It is worth noting that the de Sitter solution with time-independent $\lambda$ and $f$ was found in Ref.~\cite{Hehl4} in a wide class of Poincar\'e gravity models, but in that case invariance under parity was imposed, i.e. $g$ was set equal to zero. In our case parity breaking, $g \neq 0$, is crucial for the existence of the solution. 

For constant $f$, $g$ and $\lambda$ the three equations \eqref{all_27}, \eqref{all_28} and \eqref{all_29} become algebraic. Assuming $g \neq 0$ we have,
\begin{subequations}
\bea
\talpha (f^2 - g^2) - \al \lambda^2 - d_6 g^2 \lambda^2
+ 4 d_6 f^2 g^2 &=& 0 \;,
\label{all_5}
\\
\talpha (f+\lambda)  - d_5 g^2 \lambda +
4 d_6 g^2 f &=& 0 \;,
\label{all_6}
\\
- \talpha  + d_5 \lambda f - 2 d_6  \lambda^2 +
4 d_6 f^2  &=& 0 \;.
\label{all_7}
\eea
\end{subequations}
A useful consequence of eqs. \eqref{all_5} - \eqref{all_7} is
\be
\talpha(f^2 - g^2 - \lambda f) - 2 \al \lambda^2 = 0 \;.
\label{all_11}
\ee

Let us solve Eq.~(\ref{all_6}) for $f$:
\be
f= -\frac{\talpha -d_5g^2}{\talpha+4d_6g^2}\lambda \;.
\label{all_8}
\ee
 We substitute this into Eq.~(\ref{all_5}) and solve for $\lambda^2$. The result is
 \be
 \lambda^2 =\frac{ \talpha g^2(\talpha+ 4d_6g^2)}{(d_5^2 -4d_6^2)g^4
-[\talpha(2d_5+d_6) +4\al d_6]g^2 +\talpha (\talpha-\al)} \;.
 \label{all_9}
 \ee
Finally, we substitute $f$ from (\ref{all_8}) and $\lambda^2$ from (\ref{all_9}) into (\ref{all_11}), introduce the notation 
$$ x=g^2 \;, $$
 and obtain an
equation for $x$:
  \be
 4d_6(d_5^2-4d_6^2)x^3 -4( 2\talpha d_6^2
-4\al d_6^2 +\talpha d_5 d_6)x^2
+\talpha( \talpha d_5 -\talpha d_6 +8\al d_6)x
-\talpha^2 (\talpha-\al)=0 \;.  \label{e1}
 \ee
 This is cubic in $x$. The product of the three roots of this equation is
\be \frac{\talpha^2 (\talpha-\al)}{4d_6(d_5^2-4d_6^2)} \; ,  \label{all_85} \ee
which is positive provided that we impose the condition $d_5^2 - 4d_6^2 \equiv 4c_5(d_5 + 2d_6)>0$. The latter is equivalent, with the restriction \eqref{all_13}, to  \eqref{all_62}.
The positivity of \eqref{all_85} guarantees that there is one positive root $x_1$ for $g^2$. The two other roots are negative. Indeed, the following bilinear combination of the three roots is
$$ x_1x_2+x_1x_3+x_2x_3 = \frac{\talpha(\talpha d_5 - \talpha d_6 + 8\al d_6)}{4d_6(d_5^2-4d_6^2)} <0 $$
Furthermore, with our inequality \eqref{all_62}, we have, $$2d_5+d_6= 2(d_5+2d_6) - 3d_6 \equiv 8(c_5+16c_6)-48c_6<0 \;, $$ and hence $$\talpha(2d_5+d_6) +4\al d_6\ < 0 \;. $$
Using the latter inequality it is straightforward to see that Eq.~\eqref{all_9} gives positive $\lambda^2$ and with positive $\lambda$ we obtain a negative value for $f$ from \eqref{all_8}.

To summarize, if the condition \eqref{all_62} is satisfied together with the conditions \eqref{s1}, then there exists the self-accelerating solution \eqref{c1} with time-independent $\lambda$, $f$, $g$ and $\lambda>0$, $f<0$. The sign of $g$ can be arbitrary, since $g$ is P-odd. The value of $g^2 \equiv x$ is determined from eq.~\eqref{e1}, then the de Sitter expansion rate is given by \eqref{all_9} and $f$ is given by \eqref{all_8}. 

\section{\label{all_50}The limit of small $\lambda$}

The solution given in the previous section, although exact,  is
fairly complicated.
Having in mind the present acceleration of the Universe, we consider the limit of small $\lambda$. Let us find out the relevant corner in the parameter space of the action \eqref{all_2}. We assume the following power counting:
\[
\frac{\talpha}{|\al|} = O(\lambda^0)\;,\;f=O(\lambda^0)\;,
\]
where $|\al| \sim M_{Pl}^2$, see Eq.~\eqref{all_83}. We note in passing that this power counting does not exclude the case $\talpha << |\al|$ and/or  $f^2<<|\al|$ ; it merely means that $\lambda$ is the smallest parameter in the problem.

We make use of \eqref{all_11} to solve eqs. \eqref{all_5}, \eqref{all_6} and \eqref{all_7} for $c_5$ and $c_6$:
\begin{align*}
&c_6=\frac{\talpha\lambda (\talpha f+\al \lambda)}{16(\lambda^2-4f^2)(\talpha f^2-\talpha\lambda f - 2\al\lambda^2)} \;, 
\\
& c_5=\frac{\talpha[2\talpha f^2+\lambda f\talpha+\lambda^2(\talpha-2\al)]}{4\lambda(\lambda+2f)(f^2\talpha-\lambda f\talpha-2\al\lambda^2)}  \;.
\end{align*}
In the $\text{small-}\lambda$ limit, these equations give
\begin{align*}
& c_6=-\frac{\talpha}{64f^3}\lambda \;, \\
& c_5=\frac{\talpha}{4\lambda f} \;,
\end{align*}
or
\begin{align*}
 & \lambda = \left( -\frac{c_6\talpha^2}{c_5^3} \right)^{1/4}  \;, \\
 & f=-\frac{\talpha^{1/2}}{4(-c_5c_6)^{1/4}}  \;,
\end{align*}
so that the parameter $c_6$ must be small and $c_5$ must be large, $c_6=O(\lambda)$, $c_5=O(\lambda^{-1})$. Equation \eqref{all_11} then gives $$ g=\pm f+O(\lambda) \;.$$
As we pointed out above, the sign of $g$ can be chosen arbitrarily.

The effective cosmological constant can also be written in terms of the masses \eqref{all_80}, \eqref{all_81} of excitations about the Minkowski background:
\[
  \lambda = m\left(  \frac{m}{m_0} \right)^{1/2}\frac{(-\al)^{3/4}}{2^{1/4}(\talpha-\al)^{3/4}}  \; .
\]
At $\talpha/|\al| = O(\lambda^0)$ this shows that the small value of $\lambda$ is obtained for small mass $m$ of the spin-2 excitation, and also that large $m_0$ suppresses $\lambda$.

\section{Conclusions}
To conclude, the model \eqref{all_2} admits the self-accelerating solution,
\[
e^{\t0}_0 = 1 \;,\;\;\;\; e^{\ta }_b = e^{\lambda t} \delta^{\ta}_b
 \;,\quad
A_{\t0 \ta \tb}=f\delta_{\ta\tb}  \;, \quad
A_{\ta\tb\tc}=g\varepsilon_{\ta\tb\tc}\;,
\]
We have shown that for the most general solution with the de~Sitter metric,
 $f$ and $g$ are  necessarily  time-independent constants.  Furthermore,  we have established a direct relationship between the dark energy $\lambda$ and the mass  $m$ characteristic of the massive graviton originating from
torsion in Minkowski background. The small value of the effective cosmological constant $\lambda$ is obtained provided that there is a hierarchy between the couplings, $c_6=O(\lambda)$, $c_5=O(\lambda^{-1})$. It is worth noting that with this choice of parameters and in Minkowski background, the mass of the spin-2 state \eqref{all_80} is small, $m^2 \sim \lambda f$, while the mass of spin-0 state \eqref{all_81} is large, $m_0^2 \sim f^3/\lambda$.  In fact, in the limit of small $\lambda$, the scale $m_0$ may be above the UV cutoff of the effective low energy theory; in that case the scalar degree of freedom is absent in the spectrum about Minkowski background. We emphasize that perturbations about our self-accelerating background may have quite different properties. We plan to address this issue in a forthcoming publication.

\section*{Acknowledgements}
The authors are grateful to D.~Levkov for helpful discussions. V. Nikiforova
thanks E. Nugaev for numerous suggestions and G.~'t~Hooft for comments.
The work of V.~N. and V.~R. has been supported by Russian Science
Foundation grant 14-22-00161.


\begin{thebibliography}{20}



\bibitem{9}
  G.~R.~Dvali, G.~Gabadadze and M.~Porrati,
  ``4-D gravity on a brane in 5-D Minkowski space,''
  Phys.\ Lett.\ B {\bf 485}, 208 (2000)
  doi:10.1016/S0370-2693(00)00669-9
  [hep-th/0005016].





\bibitem{41}
C.~Deffayet, G.~R.~Dvali and G.~Gabadadze,
  ``Accelerated universe from gravity leaking to extra dimensions,''
  Phys.\ Rev.\ D {\bf 65}, 044023 (2002)
  doi:10.1103/PhysRevD.65.044023
  [astro-ph/0105068].






  \bibitem{27}
D.~Gorbunov, K.~Koyama and S.~Sibiryakov,
  ``More on ghosts in DGP model,''
  Phys.\ Rev.\ D {\bf 73}, 044016 (2006)
  doi:10.1103/PhysRevD.73.044016
  [hep-th/0512097].







\bibitem{14}
G.~Gabadadze,
  ``ICTP lectures on large extra dimensions,''
  hep-ph/0308112.


\bibitem{37}
Y. Fujii, K. Maeda, 
``The Scalar-Tensor Theory of Gravitation'' (Cambridge University Press, 2003).


  \bibitem{21}
V.~A.~Rubakov and P.~G.~Tinyakov,
  ``Infrared-modified gravities and massive gravitons,''
  Phys.\ Usp.\  {\bf 51}, 759 (2008)
  doi:10.1070/PU2008v051n08ABEH006600
  [arXiv:0802.4379 [hep-th]].
  
  
  \bibitem{36}
A.~De Felice and S.~Tsujikawa,
  ``f(R) theories,''
  Living Rev.\ Rel.\  {\bf 13}, 3 (2010)
  doi:10.12942/lrr-2010-3
  [arXiv:1002.4928 [gr-qc]].


\bibitem{39}
K.~Hinterbichler,
  ``Theoretical Aspects of Massive Gravity,''
  Rev.\ Mod.\ Phys.\  {\bf 84}, 671 (2012)
  doi:10.1103/RevModPhys.84.671
  [arXiv:1105.3735 [hep-th]].



\bibitem{35}
T.~Clifton, P.~G.~Ferreira, A.~Padilla and C.~Skordis,
  ``Modified Gravity and Cosmology,''
  Phys.\ Rept.\  {\bf 513}, 1 (2012)
  doi:10.1016/j.physrep.2012.01.001
  [arXiv:1106.2476 [astro-ph.CO]].






\bibitem{book1}
  M.~Blagojevi\'c,
  ``Gravitation and gauge symmetries,''
  Bristol, UK: IOP (2002).


\bibitem{book3}
M.~Blagojevi\'c and F.~W.~Hehl,
``Gauge Theories of Gravitation : A Reader with Commentaries''  (Imperial College Press, 2013).


\bibitem{book2}
T.~Ortin,
``Gravity and strings'' (Cambridge University Press, 2015).



\bibitem{H2}
P.~Baekler, F.~W.~Hehl and J.~M.~Nester,
``Poincare gauge theory of gravity: Friedman cosmology with even and odd parity modes. Analytic part,''
  Phys.\ Rev.\ D {\bf 83}, 024001 (2011)
  doi:10.1103/PhysRevD.83.024001
  [arXiv:1009.5112 [gr-qc]].


\bibitem{H}
  P.~Baekler and F.~W.~Hehl,
  ``Beyond Einstein-Cartan gravity: Quadratic torsion and curvature invariants with even and odd parity including all boundary terms,''
  Class.\ Quant.\ Grav.\  {\bf 28}, 215017 (2011)
  doi:10.1088/0264-9381/28/21/215017
  [arXiv:1105.3504 [gr-qc]].
  
  
 

\bibitem{Hehl4}
P.~Baekler and F.~W.~Hehl, ``A micro-deSitter
    spacetime with constant torsion: A new vacuum solution of the
    Poincar\'e gauge field theory,'' Lecture Notes in Physics
  (Springer) {\bf 176} 1--15 (1983).
\bibitem{42}
A.~V.~Minkevich,
  ``Generalized Cosmological Friedmann Equations And The De Sitter Solution,''
  Phys.\ Lett.\ A {\bf 95}, 422 (1983).
  doi:10.1016/0375-9601(83)90309-2
  

\bibitem{49} 
  K.~F.~Shie, J.~M.~Nester and H.~J.~Yo,
  ``Torsion Cosmology and the Accelerating Universe,''
  Phys.\ Rev.\ D {\bf 78}, 023522 (2008)
  doi:10.1103/PhysRevD.78.023522
  [arXiv:0805.3834 [gr-qc]].
  

\bibitem{44}
A.~V.~Minkevich,
  ``De Sitter spacetime with torsion as physical spacetime in the vacuum,''
  Mod.\ Phys.\ Lett.\ A {\bf 26}, 259 (2011)
  doi:10.1142/S0217732311034797
  [arXiv:1002.0538 [gr-qc]].

  
\bibitem{50}
  X.~C.~Ao and X.~Z.~Li,
  ``Torsion Cosmology of Poincar\'e gauge theory and the constraints of its parameters via SNeIa data,''
  JCAP {\bf 1202} (2012) 003
  doi:10.1088/1475-7516/2012/02/003
  [arXiv:1111.2385 [gr-qc]].
    


\bibitem{47}
  G.~Chee and Y.~Guo,
  ``Exact de Sitter solutions in quadratic gravitation with torsion,''
  Class.\ Quant.\ Grav.\  {\bf 29} (2012) 235022
  doi:10.1088/0264-9381/29/23/235022
  [arXiv:1205.5419 [gr-qc]].
    

\bibitem{51}
 C.~Q.~Geng, C.~C.~Lee and H.~H.~Tseng,
  ``Scalar-Torsion Cosmology in the Poincar\'e Gauge Theory of Gravity,''
  JCAP {\bf 1211} (2012) 013
  doi:10.1088/1475-7516/2012/11/013
  [arXiv:1207.0579 [gr-qc]].
 


\bibitem{46}
A.~S.~Garkun, V.~I.~Kudin and A.~V.~Minkevich,
  ``To theory of asymptotically stable accelerating Universe in Riemann-Cartan spacetime,''
  JCAP {\bf 1412}, no. 12, 027 (2014)
  doi:10.1088/1475-7516/2014/12/027
  [arXiv:1410.0460 [gr-qc]].


\bibitem{48}
  J.~Lu and G.~Chee,
  ``Cosmology in Poincar\'e gauge gravity with a pseudoscalar torsion,''
  JHEP {\bf 1605} (2016) 024
  doi:10.1007/JHEP05(2016)024
  [arXiv:1601.03943 [gr-qc]].







\bibitem{29}
K.~Hayashi and T.~Shirafuji,
  ``Gravity from Poincar\'e Gauge Theory of the Fundamental Particles. 1. Linear and Quadratic Lagrangians,''
  Prog.\ Theor.\ Phys.\  {\bf 64}, 866 (1980)
  Erratum: [Prog.\ Theor.\ Phys.\  {\bf 65}, 2079 (1981)].
  doi:10.1143/PTP.64.866

\bibitem{30}
K.~Hayashi and T.~Shirafuji,
  ``Gravity From Poincar\'e Gauge Theory of the Fundamental Particles. 3. Weak Field Approximation,''
  Prog.\ Theor.\ Phys.\  {\bf 64}, 1435 (1980)
  Erratum: [Prog.\ Theor.\ Phys.\  {\bf 66}, 741 (1981)].
  doi:10.1143/PTP.64.1435

\bibitem{31}
K.~Hayashi and T.~Shirafuji,
  ``Gravity From Poincar\'e Gauge Theory of the Fundamental Particles. 4. Mass and Energy of Particle Spectrum,''
  Prog.\ Theor.\ Phys.\  {\bf 64}, 2222 (1980).
  doi:10.1143/PTP.64.2222

\bibitem{32}
E.~Sezgin and P.~van Nieuwenhuizen,
  ``New Ghost Free Gravity Lagrangians with Propagating Torsion,''
  Phys.\ Rev.\ D {\bf 21}, 3269 (1980).
  doi:10.1103/PhysRevD.21.3269
  
  



\bibitem{45}
V.~P.~Nair, S.~Randjbar-Daemi and V.~Rubakov,
  ``Massive Spin-2 fields of Geometric Origin in Curved Spacetimes,''
  Phys.\ Rev.\ D {\bf 80}, 104031 (2009)
  doi:10.1103/PhysRevD.80.104031
  [arXiv:0811.3781 [hep-th]].

\bibitem{33}
V.~Nikiforova, S.~Randjbar-Daemi and V.~Rubakov,
  ``Infrared Modified Gravity with Dynamical Torsion,''
  Phys.\ Rev.\ D {\bf 80}, 124050 (2009)
  doi:10.1103/PhysRevD.80.124050
  [arXiv:0905.3732 [hep-th]].


\bibitem{34}
C.~Deffayet and S.~Randjbar-Daemi,
  ``Non linear Fierz-Pauli theory from torsion and bigravity,''
  Phys.\ Rev.\ D {\bf 84}, 044053 (2011)
  doi:10.1103/PhysRevD.84.044053
  [arXiv:1103.2671 [hep-th]].
  
  
  
  
  \bibitem{Holst}
  S.~Holst,
  ``Barbero's Hamiltonian derived from a generalized Hilbert-Palatini action,''
  Phys.\ Rev.\ D {\bf 53}, 5966 (1996)
  doi:10.1103/PhysRevD.53.5966
  [gr-qc/9511026].



\bibitem{Hehl3}
F.~W.~Hehl, P.~Von Der Heyde, G.~D.~Kerlick and J.~M.~Nester,
  ``General Relativity with Spin and Torsion: Foundations and Prospects,''
  Rev.\ Mod.\ Phys.\  {\bf 48}, 393 (1976).
  doi:10.1103/RevModPhys.48.393

\bibitem{obs_constraints}
W.~T.~Ni,
  ``Searches for the role of spin and polarization in gravity,''
  Rept.\ Prog.\ Phys.\  {\bf 73}, 056901 (2010)
  doi:10.1088/0034-4885/73/5/056901
  [arXiv:0912.5057 [gr-qc]].
  






  


\end{thebibliography}
\end{document}